\documentclass[preprint,tightenlines,eqsecnum,floats,aps,amsmath,amssymb,nofootinbib,prd,showpacs]{revtex4}

\usepackage{amssymb}
\usepackage{stmaryrd}
\usepackage{amsmath}
\usepackage{amsfonts}
\usepackage{mathrsfs}
\usepackage{CJK}
\usepackage{amsmath,amssymb,amsfonts}
\usepackage{graphicx}
\usepackage{subfigure}
\usepackage[colorlinks,linkcolor=blue,anchorcolor=red,citecolor=green]{hyperref}
\def\be{\begin{equation}}
\def\ee{\end{equation}}
\def\ba{\begin{eqnarray}}
\def\ea{\end{eqnarray}}
\def\nn{\nonumber}
\def\dt{\tilde{\delta}}

\newcommand{\eqnref}[1]{(\ref{#1})}

\begin{document}

\title{Collision of spinning particles near BTZ black holes}

\author{ Xulong Yuan, Yunlong Liu}
\affiliation{School of Physics, South China University of
Technology, Guangzhou 510641, P.R. China}

\author{ Xiangdong Zhang\footnote{Corresponding author. scxdzhang@scut.edu.cn}}
\affiliation{School of Physics, South China University of
Technology, Guangzhou 510641, P.R. China}

\date{\today}


\begin{abstract}
We study the collision property of spinning particles near a Ba\~nados-Teitelboim-Zanelli
(BTZ) black hole. Our results show that although the center-of-mass energy of two ingoing particles diverges if one of the particles possesses a critical angular momentum, however, particle with critical angular momentum can not exist outside of the horizon due to the violation of timelike constraint. Further detailed investigation indicates that only a particle with a subcritical angular momentum is allowed to exist near an extremal rotating BTZ black hole and the corresponding collision  center-of-mass energy can be arbitrarily large in a critical angular momentum limit.
\end{abstract}

\

\maketitle

\section{Introduction}
The particle collision near a black hole background has long history. The possibility of having infinite center-of-mass energy collision near a black hole was first pointed out by Piran, Shaham, and Katz in 1975\cite{Piran1975}. In 2009, Ba\~nados, Silk, and West\cite{banados2009}, rediscovered this mechanism, known as the BSW process, and they pointed out that because of infinite center-of-mass energy caused by collision, the rotating black holes can thus act as particle
accelerators\cite{banados2009,banados2011}. Along this line, many aspects of BSW mechanism with various black hole backgrounds have been investigated. For instance, the Kerr naked singularity\cite{Patil2011}, the charged spinning black hole\cite{weishw2010}, the Kerr-(anti)de-Sitter black hole spacetime \cite{liyang2011} and the universal property of rotating black holes was given in \cite{zaslav2010}. Other research related to higher or lower dimensional spacetime background \cite{Abdujabbarov2013,Tursunov2013,Sadeghi2014} is also interesting, such as five dimensional Kerr black hole can be found in\cite{Abdujabbarov2013} and three dimensional rotating charged hairy black hole have been studied in \cite{Sadeghi2014}. Furthermore, the BSW mechanism can help us to optimize the collisional penrose process which extracts energy from a black hole through particle collision\cite{piran1977,M.Bejger2012,Schnittman2014,berti2015,Leiderschneider,Ogasawara2016,Harada2016,Ogasawara2017,K.i.Maeda18}.

In three dimensional spacetime, there exists a
typical stationary black hole solution with a negative cosmological constant which was first discovered by Ba\~nados-Teitelboim-Zanelli (BTZ)\cite{btz1992}. This black hole solution, because of its similarity and simplicity compared with the (3+1)-dimensional Kerr black hole, has received increasing attention recently. For example, the spinless particles collision around
BTZ black hole has been study in \cite{tsukamoto2017}. People are interested in the (2+1)-dimensional BTZ black hole because it can
be a good toy model which help to gain more deep understanding of the same problem in the Kerr spacetime, since in the BTZ background the analytical expression usually is possible\cite{ Tsukamoto2019,Mann2009,Rocha2011}, while in the Kerr spacetime the same analytical treatment for the same problem is generally very difficult. For example, the collision
of fast rotating dust thin shells in (2+1)-dimension is much simpler compared with the (3+1)-dimensional Kerr spacetime\cite{Tsukamoto2019,Mann2009,Rocha2011}.

On the other hand, many authors focus on  point particle whose trajectory is a geodesic. However a real particle should be an extended body with
inclusion of self-interaction. Compared with the spinless particle, the orbit
of a spinning test particle is no longer a geodesic, and it has been shown that\cite{papapetrou1951,dixon1970,wald1972,kerr1963,gaosj2016,banados2016,zhangyp2016} the equations of motion of spinning particles
around a given spacetime background is discribed by the
Mathisson-Papapetrou-Dixon(MPD) equations \cite{saijo1998,hackmann2014,Jefremov2015}.
By collecting these results, the authors in \cite{gaosj2016} show that the collision center-of-mass energy could be divergent for extremal Kerr black hole. With these motivations, our research in this paper is devoted to study the collision of spinning particles around the BTZ black hole.

The paper is organized as follows: In section \ref{Equation}, we introduce Mathission-Papapetrou-Dixon(MPD) equations, which describes  the spinning particles' motion in curved spacetime, and apply it to Ba\~nados-Teitelboim-Zanelli(BTZ) black hole. In section \ref{CM energy}, we obtain the collision center-of-mass energy of spinning particles and find the condition for the divergence of center-of-mass energy with either of the particle possessing critical total angular momentum. Then in section \ref{motion}, the motion of spinning particles with critical and subcritical angular momentum near the event horizon are analyzed in details, and it's shown a spinning particle with subcritical total angular momentum are allowed to exist on or outside the horizon. Next, in section \ref{COLU}, collision of two spinning particles with subcritical total angular momenta near the horizon are calculated, and the diverging center-of-mass energy in the critical limit is obtained. Conclusions are given in the last section \ref{conclusion}. Through out the paper, we adopt the convention that the speed of light $c=1$.

\section{Equations of motion for spinning particles}\label{Equation}

The the metric of BTZ black hole  in the Boyer-Lindquist coordinates  reads\cite{btz1992}

\begin{eqnarray}
\mathrm{d}s^2&=&-g(r)\mathrm{d}t^2+{\mathrm{d}r^2\over g(r)}+r^2\left(\mathrm{d}\phi-{r_{+}r_{-}\over lr^2}\mathrm{d}t\right)^2,\label{2.26}
\end{eqnarray}
where
\begin{eqnarray}
g(r)&=&{(r^2-r_{+}^2)(r^2-r_{-}^2)\over l^2r^2},\label{2.27}\\
\ea
and
\ba
M&=&{r_{+}^2+r_{-}^2\over 8Gl^2},\label{2.28}\\
J&=&{r_{+}r_{-}\over 4Gl}.\label{3.9}
\end{eqnarray}
and  $r=r_{+}$ is the outer horizon, $r=r_{-}$ is the inner horizon,  $M$ is the ADM mass, $J$ is the angular momentum and $l$ is a parameter determined by the negative cosmological constant $\Lambda$ ($l^2=-\Lambda/3$).
Note that for the angular momentum $J$, ${|J| }\leq M l$ must be satisfied. When the black hole is extremal($r_{+}= r_{-}$),   we have $|J| = M l$.

Under the given BTZ spacetime, the spinning particle's motion can be described by MPD equations\cite{wald1972,gaosj2016}
\begin{eqnarray}\label{mpd}
&\frac{D}{D\tau }P^{a}=-\frac{1}{2}R_{bcd}^{a} v ^{b}S^{cd},\label{mpd1}\\
&\frac{D}{D\tau }S^{ab}=P^{a} v^{b}-P^{b} v^{a}, \label{mpd2}
\end{eqnarray}
Along the center-of-mass world line $z(\tau)$, $\upsilon ^{a}=(\frac{\partial }{\partial \tau })^{a}$
is the tangent vector,  $\frac{D}{D\tau }$ is the covariant derivative,  $P^{a}$ is the momentum of the spinning particles, and $S^{ab}$ is the spinning angular momentum tensor.

In order to obtain the detailed relation between momentum $P^a$ and $v^a$ , supplementary conditions are required to be imposed: \cite{banados2016,gaosj2016}
\begin{eqnarray}
\label{fixcenterofmass} &&S^{ab}P_{b}=0,\\
\label{normalizetau} &&P^a v_a=-m ,
\end{eqnarray}
where $\tau$ is not necessarily the proper time of the spinning particle. Combining the Eqs. \eqref{mpd}, \eqref{fixcenterofmass} and \eqref{normalizetau}, the difference between $v^{a}$ and $u^{a}$reads\cite{saijo1998,gaosj2016}
\be\label{vaua}
mv^{a}-P^{a}=\frac{S^{ab}R_{bcde}P^{c}S^{de}}{2(m^{2}+\frac{1}{4}R_{bcde}S^{bc}S^{de})}.
\ee
With direct calculation , we find that $v^a=u^a$ in BTZ spacetime where $u^a\equiv {P^a / m}$. It should be note that the velocity $v^a$ is parallel to the momentum $u^a$ in the specific property in (2+1)-dimensions, and of course in generally not valid in 4-dimensional spacetime.

Note that, there are two Killing vector fields $\xi ^a=\left(\partial/\partial t\right)^a$ and $\phi^a=\left(\partial/\partial \phi\right)^a$ in the BTZ spacetime because  BTZ spacetime is  axi-symmetric and stationary.
They can be expanded in the  orthonormal triad basis ${e_{a}^{(\nu)}}$ as
\ba
\xi_a&=&-\sqrt{f(r)}e_{a}^{(0)}-{r_{+}r_{-}\over lr}e_{a}^{(2)},\nn\\
\phi_a&=&r e_{a}^{(2)}\label{3.9}.
\ea
where
\ba
e_{a}^{(0)}&=&\sqrt{f(r)}(\mathrm{d}t)_a,  \label{3.6}\nn\\
e_{a}^{(1)}&=&{1\over \sqrt{f(r)}}(\mathrm{d}r)_a,\label{3.7}\nn \\
e_{a}^{(2)}&=&r\left((\mathrm{d}\phi)_a-{r_{+}r_{-}\over lr^2}(\mathrm{d}t)_a\right).\label{3.8}
\ea

Then, a corresponding conserved quantity can be defined by  Killing vector field $\xi^{a}$ as follow:
\begin{equation}
Q_{\xi }=P^{a}\xi _{a}-\frac{1}{2}S^{ab}\triangledown_{b}\xi_{a} \label{killingcons}
\end{equation}

From the equation above, two conserved quantities can be obtained, namely the energy of per unit mass of the particle $E_m$, and the angular momentum per unit mass of the particle $J_m$, they are
\ba
E_m&=&-u^a\xi _a+\frac{1}{2m}S^{ab}\nabla _b\xi _a,\nn\\
J_m&=&u^a\phi _a-\frac{1}{2m}S^{ab}\nabla _b\phi _a. \label{emjmcons}
\ea

Combining with Eq.\eqref{mpd2} and \eqref{fixcenterofmass} we can introduce the spin $s$ of the particle as
\begin{equation}
s^2:=\frac{1}{2m^2}S^{ab}S_{ab},\label{spinconserved}
\end{equation}
where $s$ is the spin of unit mass.
What's more, Combining  with Eq.\eqref{mpd1}, \eqref{mpd2} and \eqref{fixcenterofmass}, the spin tensor can be written reversely as
\begin{equation}\label{spintensor}
S^{(a)(b)}=-m\varepsilon ^{(a)(b)}_{~~~~~~(c)}u^{(c)}s,
\end{equation}
where $\varepsilon _{(a)(b)(c)}$ is the completely anti-symmetric tensor with the component $\varepsilon _{(a)(b)(c)}=1$.

From Eq.(\ref{spintensor}), the non-zero components of the spin tensor can be
expressed in terms of $u^{(a)}$ as
\ba\label{spintcomp}
S^{(0)(1)}&=&-msu^{(2)},\nn\\
S^{(0)(2)}&=&msu^{(1)},\nn\\
S^{(1)(2)}&=&msu^{(0)}.
\ea

The explicit expressions of the energy and
the angular momentum per unit mass $E_\text{m}$ and $J_\text{m}$ in terms of $u^{(a)}$ can be obtained by using Eq. (\ref{spintcomp}) and Eq. (\ref{emjmcons}) as:
\ba
E_m&=&\sqrt{f(r)}u^{(0)}+\left({r_{+}r_{-}\over lr}+{rs\over l^2}\right)u^{(2)}, \label{emcomp}\\
J_m&=&s\sqrt{f(r)}u^{(0)}+\left({r_{+}r_{-}s\over lr}+r\right)u^{(2)}. \label{jmcomp}
\ea
Solving Eq. (\ref{emcomp}) and Eq. (\ref{jmcomp}) gives
\ba
u^{(0)}&=&\frac{l \left(l E_m \left(l r^2+r_- r_+ s\right)-J_m \left(l r_- r_++r^2 s\right)\right)}{r \sqrt{\left(r^2-r_-^2\right) \left(r^2-r_+^2\right)} (l^2-s^2)},\\
u^{(2)}&=&\frac{J_m-E_m s}{r-\frac{r s^2}{l^2}}.
\ea
By considering the normalization condition of momentum $u^{(a)}u_{(a)}=-m^2$ , we obtain the $u^{(1)}$ as follow:
\be
(u^{(1)})^2=(u^{(0)})^{2}-(u^{(2)})^2-m^2. \label{3.23}
\ee

For direct comparison to the spinless case in \cite{tsukamoto2017}, now we express the momentum in coordinate basis:
\ba
p^t(r)&=&{\mathrm{d}t\over\mathrm{d}\tau}={W(r)\over f(r)}, \label{ptr}\\
p^r(r)&=&{\mathrm{d}r\over\mathrm{d}\tau}=\rho\sqrt{Y(r)}, \label{3.19}\\
p^\phi(r)&=&{\mathrm{d}\phi\over\mathrm{d}\tau}={r_{+}r_{-}W(r)\over lf(r)r^2}+{l^2(J_m-E_ms)\over r^2(l^2-s^2)}. \label{3.25}
\ea
where
\ba
W(r)&=&\frac{E_\text{m} l \left(l r^2+r_{-} r_{+} s\right)-J_\text{m} \left(l r_{-} r_{+}+r^2 s\right)}{r^2 \left(l^2-s^2\right)},
\label{srdef}\\
Y(r)&=&W^2(r)-\left(m^2+\left({J_\text{m}-E_\text{m}s\over r(1-{s^2\over l^2})}\right)^2\right)f(r), \label{rrdef}\\
\rho&=&+1\text{ for outwards direction, }-1\text{ for inwards direction.}\nn
\ea
We define the critical angular momentum as
\ba
J_c\equiv{E_ml(lr_{+}+r_{-}s)\over lr_{-}+r_{+}s},\label{critij}
\ea
and a particle with critical angular momentum $J_c$ corresponds to:
\ba
W_i(r_{+})=0,
\ea where $i=1,2$ refers to particle 1 or particle 2 in the collision process.
When the particle's spin $s=0$, the critical angular momentum introduced here will be reduced to the spinless case, which already has been investigated in \cite{tsukamoto2017}.

The timelike constraint of Eq. (\ref{ptr}) means $p^t(r)>0$ outside the horizon for massive particles, which in turn implies $W_i(r)>0$. Therefore, for particles with the angular momentum $J_m\leq J_c$, the positivity of $W_i(r)$ gives rise to a constraint on particle's spin as $l^2-s^2>0$. Therefore in the following sections, we restrict ourselves to the case $-l<s<l$.

\section{center-of-mass energy of the collision}\label{CM energy}
In this section we intend to find the condition required for infinite center-of-mass energy collision of two spinning massive particles near the BTZ horizon. They are particle $i=1,2$ that starts at infinity with masses $m_i$, energy per unit mass $E_{mi}$, total angular momenta per unit mass $J_{mi}$, and spins $s_i$ respectively, falling to the black hole and colliding near the event horizon. Then the collision center-of-mass energy $E_{cm}$ is given by\cite{tsukamoto2017,gaosj2016}:
\ba
E^2_{cm}&\equiv&-(p^\mu_1(r)+p^\mu_2(r))(p_{1\mu}(r)+p_{2\mu}(r))=m_1^2+m_2^2 \nn\\
&&+{W_1(r)W_2(r)-\sqrt{Y_1(r)Y_2(r)}\over f(r)}-2{l^4(J_{m1}-E_{m1}s_1)(J_{m2}-E_{m1}s_2)\over r^2(l^2-s_1^2)(l^2-s_2^2)},\label{4.1}
\ea
where $Y_i(r)$ and $W_i(r)$ are defined by Eq. (\ref{srdef}) and (\ref{rrdef}) with $i=1,2$ again labeling particle 1 or particle 2.

We find that the third term of Eq. \eqref{4.1} is a ${0\over 0}$ type when $r$ approaches to the event horizon $r_{+}$, so we first need to regularize this term as
\ba
\lim_{r\rightarrow r_{+}}2{W_1(r)W_2(r)-\sqrt{Y_1(r)Y_2(r)}\over f(r)}
={W_2(r_{+})\over W_1(r_{+})}Z_1+{W_1(r_{+})\over W_2(r_{+})}Z_2,\\
\text{in which }Z_i=\left(m_i^2+\left({J_{mi}-E_{mi}s_i\over r(1-{s_i^2\over l^2})}\right)^2\right)>0.
\ea
It is easily to see that $E^2_{cm}$ blows up with $r\rightarrow r_{+}$ if one of the particle has the critical angular momentum $J_c$( which means $W_i(r_{+})=0$). If both particles possess $J_c$, then we have,
\ba
{W_2(r_{+})\over W_1(r_{+})}={W'_2(r_{+})\over W'_1(r_{+})}=\frac{E_{m1}( l r_-+ r_+ s_2)}{E_{m2} (l r_-+ r_+ s_1)},
\ea
in which $'$ denotes derivative with respect to $r$. For equal spin collision ($s_1=s_2$), the ratio ${W_2(r_{+})\over W_1(r_{+})}={E_{m1}\over E_{m2}}$ becomes finite value, which is similar to the spinless case \cite{tsukamoto2017}. Therefore, the only possibility for the center-of-mass energy goes to infinity is one of the spin, for example, $s_1$, satisfies
\ba
s_1=s_c=-{lr_{-}\over r_{+}}.
\ea
However, this is equivalent to require $J_{m1}=J_{c1}$ to be infinity according to Eq. (\ref{critij}) and thus is impossible to achieve in practice.

\section{motion of a particle with critical and subcritical total angular momentum}\label{motion}
In last section, we showed that if one of the collision particle possesses critical angular momentum, the center-of-mass energy $E_{cm}$ will blow-up. However, to solid this conclusion, we still need to check whether the particle with critical angular momentum $J_c$ can satisfy other constraints such as timelike constraint in subsection \ref{superlu} and  radial equation of motion which guarantees the particles can reach the horizon. Therefore, the aim of this section is to discuss these constraints carefully.

First we note that for spinless case\cite{tsukamoto2017}, a particle with critical total angular momentum $J_m=J_c$ is not allowed to exist outside the event horizon while one with subcritical angular momentum  can be allowed. Later we shall investigate the same issue by taking account of spin effect in subsections \ref{radial} to \ref{subc}.
\subsection{Timelike constraint of $p^t(r)$}\label{superlu}
The first subsection is devoted to the timelike constraint of $p^t(r)$. To avoid superluminality, $p^t(r)$ should be non-negative, from Eq. (\ref{ptr}) we have,
\ba
p^t(r)={W(r)\over f(r)}\geq0,
\ea
since $f(r)>0$, the above equation is equivalent to,
\ba
W(r)=\frac{l \left(l E_m \left(l r^2+r_- r_+ s\right)-J_m \left(l r_- r_++r^2 s\right)\right)}{r  (l^2-s^2)}\geq 0.\label{srtl}
\ea
Eq. \eqnref{srtl} gives a restriction of $J_m$ to ensure $p^t(r)\geq 0$ near the event horizon where the infinite center-of-mass energy collision takes place. Considering extremal black hole and the case $-l<s<l$, this gives rise to
\ba
J_m\leq E_m l=J_c,\label{jctl}
\ea
which means for $J_m<J_c$, the timelike condition is satisfied. However, for massive particle, when the total angular momentum takes critical value $J_m=J_c$, the timelike condition is violated. Therefore, in the following sections, we will consider the subcritical total angular momentum $J_m<J_c$.

\subsection{Radial motion of the particle: $Y(r)$}\label{radial}
Now we come to the radial motion of the particle , we start with the expression of $p^r(r)$ \eqref{3.19}, and obtain the radial equation of motion of the spinning particle:
\ba
{1\over 2}{p^r(r)}^2+V(r)=0,\label{radialv}
\ea
where $V(r)$ is the radial effective potential defined by $V(r)\equiv -Y(r)/2$, and $\tau$ is the geodesic parameter. Particles are only allowed to exist in regions where $V(r)\leq 0$ or $Y(r)\geq 0$ from Eq. \eqref{radialv}. 

For massive particle with $m\ne 0$ we consider, the tendency of $Y(r)$ at infinity is
\ba
\lim_{r\rightarrow\infty}Y(r)=-m^2\times\infty<0\label{limr}
\ea
which implies a massive particle cannot escape to infinity. 

Since the expression of $Y(r)$ for massive particle is complicated, we will investigate it with subcritical total angular momentum in \ref{subc}, especially for extremal black hole.

\subsection{Motion of a particle with the subcritical total angular momentum}\label{subc}
In last subsection, we already know particles with critical total angular momentum cannot exist outside the event horizon. So we consider a particle with subcritical total angular momentum $J_m$($J_m\leq J_c$ for the case $-l<s<l$ according to Eq. \eqref{jctl}):
\ba
J_m\equiv J_c-\dt={E_ml(lr_{+}+r_{-}s)\over lr_{-}+r_{+}s}-\dt,\label{subcritj}
\ea
and try to find the range of $\dt$ that enables the particle to exist outside the black hole (i.e. $Y(r)\geq 0$). With this well-defined subcritical total angular momentum, we obtain the corresponding function $Y(r)$ as follows:

\ba
Y(r)&=&Y_c(r)+\left\{\dt(-2E_ml(r^2-r_{+}^2)(l^2-s^2)(lr_{+}+r_{-}s)-(lr_{-}+r_{+}s)\right.\nn\\
&&\left.(l^2(-r^2+r_{-}^2+r_{+}^2)+2lr_{-}r_{+}s+r^2s^2)\dt)\right\}/\left(r^2(l^2-s^2)^2(lr_{-}+r_{+}s)\right).\label{rrrcrit}
\ea
where
\ba
Y_c(r)=-{(r^2-r_{+}^2)\over r^2}\left({E_m^2l^2(r_{+}^2-r_{-}^2)\over (lr_{-}+r_{+}s)^2}+{m^2(r^2-r_{-})^2\over l^2}\right).\label{critir}
\ea
 When spin $s$ is taken as zero, our result will reduce to the spinless case by identifying $\dt=\frac{r_+l}{r_-}\delta$ with $\delta$ introduced in \cite{tsukamoto2017}.
From Eq. (\ref{rrrcrit}), a particle with subcritical total angular momentum can exist on the event horizon or nearby outside of the black hole since
\ba
Y(r_{+})=\left({(lr_{-}+r_{+}s)\dt\over r_{+}(l^2-s^2)}\right)^2>0.
\ea

Now we come to discuss $Y'(r)$, which is the derivative of $Y(r)$ with respect to $r$, determines how much the collision point departures from the event horizon $r_+$\cite{tsukamoto2017}.
First, $Y'(r)$ with critical total angular momentum( i.e. $\dt=0$ ) is
\ba
Y_c'(r_+)=\frac{2 \left(r_-^2-r_+^2\right) \left(l^4 E_m^2+m^2 \left(l r_-+r_+ s\right){}^2\right)}{l^2 r_+ \left(l r_-+r_+ s\right){}^2}\label{criticalR'}
\ea
 on the event horizon. For the extremal case with critical total angular momentum, since we have $r_+=r_-$, by using Eqs. (\ref{critir}) and (\ref{criticalR'}), we obtain $Y(r_+)=Y'(r_+)=0$ on the event horizon.

Then with subcritical total angular momentum, $Y'(r)$ is relevant to $\dt$
\ba
Y'(r)=Y'_c(r)-{2l\dt(-2E_mr_{+}^2(l^2-s^2)(lr_{+}+r_{-}s)+(lr_{-}+r_{+}s)(l(r_{-}^2+r_{+}^2)+2r_{-}r_{+}s)\dt)\over r^3(lr_{-}+r_{+}s)(l^2-s^2)^2},\nn\\
\ea
On the event horizon the above equation becomes
\ba
&Y'(r_{+})=D_2\dt^2+D_1\dt+D_0,\label{R'onr+}
\ea
where the coefficients read
\ba
D_2&=&-\frac{2 l \left(l \left(r_-^2+r_+^2\right)+2 r_- r_+ s\right)}{r_+^3 \left(l^2-s^2\right)^2},\\
D_1&=&\frac{4 l E_m \left(l r_++r_- s\right)}{r_+ \left(l^2-s^2\right) \left(l r_-+r_+ s\right)},\\
D_0&=&\frac{2 \left(r_-^2-r_+^2\right) \left(l^4 E_m^2+m^2 \left(l r_-+r_+ s\right){}^2\right)}{l^2 r_+ \left(l r_-+r_+ s\right){}^2}.
\ea

The coefficient $D_2<0$ by considering $s^2< l^2$, the sign of $D_2$ is the same as in spinless case\cite{tsukamoto2017}. Along the same line of \cite{tsukamoto2017}, by solving \eqref{R'onr+}, we found there exists a $E_\text{max}$
\ba
E_\text{max}=\frac{m \sqrt{r_+^2-r_-^2} \sqrt{l r_-^2+l r_+^2+2 r_+ r_- s}}{l^{3/2} r_-},\label{5.18}
\ea if the unit mass energy $E_m<E_\text{max}$, the corresponding
$Y'(r_{+})$ is always negative; on the contrary if $E_m\geq E_\text{max}$ which is more interesting, the corresponding  $Y'(r_{+})$ can be non-negative in the range of $\dt_L\leq\dt\leq\dt_R$ and is negative elsewhere with the boundaries defined as
\ba
\dt_L&=&\frac{l^3 E_m r_+^2 (l^2-s^2) \left(l r_++r_- s\right)-\sqrt{\Delta}}{l^3 \left(l r_-+r_+ s\right) \left(l \left(r_-^2+r_+^2\right)+2 r_- r_+ s\right)},\nn\\
\dt_R&=&\frac{l^3 E_m r_+^2 (l^2-s^2) \left(l r_++r_- s\right)+\sqrt{\Delta}}{l^3 \left(l r_-+r_+ s\right) \left(l \left(r_-^2+r_+^2\right)+2 r_- r_+ s\right)},\label{deltaboun}\\
\Delta&=&l^3 r_+^2 (l^2-s^2)^2 \left(l r_-+r_+ s\right){}^2 \nn\\
&&\left(l^3 E_m^2 r_-^2+m^2 \left(r_-^2-r_+^2\right) \left(l \left(r_-^2+r_+^2\right)+2 r_- r_+ s\right)\right).\nn
\ea

For the extremal black hole, $E_\text{max}=0$. It worths to note that particles with subcritical total angular momenta $J_c-\dt$ satisfying $\dt_L\leq\dt\leq\dt_R$, have $Y(r_+)>0$ with $Y'(r_+)>0$ and luckily, can exist outside the event horizon, which is in contrast to the non-existence of particles with critical total angular momentum in subsection \ref{superlu}.

For spinless particle, the infinite center-of-mass energy collision happens at the extreme point of $Y(r)$ where usually serves as return point of particle. This is because in BTZ spacetime, except the point where the particle starts to fall, $Y(r)$ has no other zero point which is usually taken as the collision point\cite{tsukamoto2017}. In order to find this turning point of radial motion, we solve the equation $Y'(r)=0$ with the positive root $r_m$:
\ba\label{rmrp}
r_m=r_{+}\left(1+{Y'(r_{+})l^2\over 2r_{+}m^2}\right)^{1\over 4}.
\ea
Consequently whether $r_m$ is greater than event horizon $r_+$ relies directly on the sign of  $Y'(r_+)$ that has been analysed above. It is shown
when $E_m\geq E_\text{max}$ and $\dt_L\leq\dt\leq\dt_R$ satisfied, we have $Y'(r_{+})\geq 0$ which in turn implies
\ba
r_m\geq r_{+},\label{rm}
\ea
Eq.\eqref{rm} means the turning point of radial motion is on or outside the event horizon.

After applying the extremal condition $r_{-}=r_{+}$, $E_\text{max}=0$,  the boundaries \eqref{deltaboun} become:
\ba
\dt_L&=&\frac{1}{2} E_m \left(l-s-\sqrt{(l-s)^2}\right),\nn\\
\dt_R&=&\frac{1}{2} E_m \left(l-s+\sqrt{(l-s)^2}\right).
\ea
Thus, the relation between $r_m$ and $r_{+}$ can be summarized as below:

Since $-l<s<l$, we have $\dt_L=0\leq\dt\leq E_m(l-s)=\dt_R$ or equivalently $E_ms\leq J_m\leq E_m l$, $r_m\geq r_{+}$;  for other values of $\dt$,
$r_m<r_{+}$ which is disfavored by the current discussion.

\begin{figure}[htp]
\includegraphics [width=0.7\textwidth]{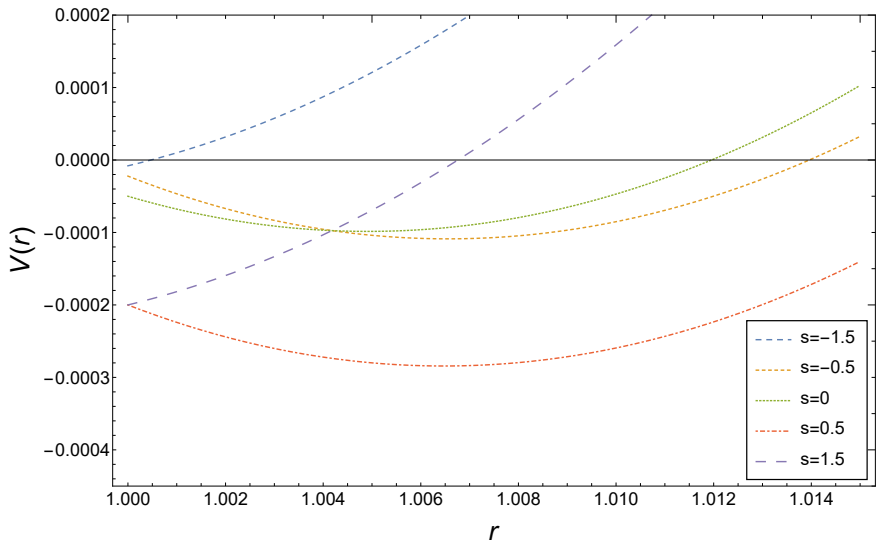}
\caption{ \label{5} The examples of effective potential of radial motion $V(r)=-{1\over 2}Y(r)$ of a particle with different spins $s$ and a subcritical total angular momentum $J_m=E_ml-\dt<J_c$ in an extremal BTZ spacetime. The minimum point of $V(r)$: $r_m$ with spin $s=-0.5,0,0.5$ are greater than $r_{+}$. Here $r_{-}=r_{+}=E_m=l=m=1,\dt=0.01$, and the longitudinal axis marks the event horizon $r_{+}$.}
\label{Fig.1}
\end{figure}

In Fig. \ref{Fig.1}, $\dt$ has been taken as 0.01 and we compare the effective potentials of radial motion $V(r)=-Y(r)/2$ of a particle with different spins $s$ and a subcritical total angular momentum $J_m=E_ml-\dt$ in an extremal BTZ spacetime, in which the minimum points mark $r_m$ where the particle is about to return, and it is shown $r_m$ with spins satisfying $-l<s<l$ are greater than $r_{+}$.

  \section{collision of spinning particles near an extremal BTZ black hole and its critical total angular momentum limit}\label{COLU}
Note that the divergence condition for center-of-mass energy with critical total angular momentum $J_c$ in section \ref{CM energy} was found to be unavailable in subsection \ref{superlu}. Due to the timelike constraint, both spinning particles are required to possess subcritical values of total angular momentum $J_{m1}=E_{m1}l-\dt_1,J_{m2}\leq E_{m2}l$, with $\dt_{L1}\leq\dt_1\leq\dt_{R1}$ because we pick $r_{m1}$ as the collision point. Then  we consider the collision center-of-mass energy $E^2_{cm}$ by taking the limit $\dt_1\rightarrow 0$.


\ba
\lim_{\dt_1\rightarrow 0}E_{cm}^2(r_{m})=m_1^2+m_2^2+Q-{2l^4(E_{m1}(l-s_1))J_{m2}\over r_+^2(l^2-s_1^2)(l^2-s_2^2)},
\ea
in which
\ba
Q\equiv\lim_{r\rightarrow r_{m}}2{W_1(r_{m})W_2(r_{m})-\sqrt{Y_1(r_{m})Y_2(r_m)}\over f(r_m)}.
\ea
Both numerator and denominator of $Q$ tend to zero in the limit of $\dt_1\rightarrow 0$. For this reason, we express $Q$ using L'Hopital's rule with respect to $\dt_1$ as
\ba
Q&=&\lim_{\dt_1\rightarrow 0}{W_2(r_m)\over \dot{f}(r_m)}\left(2\dot{W}_1(r_m)-{\dot{Y}_1(r_m)\over \sqrt{Y_1(r_m)}}\right)\nn\\
&=&\lim_{\dt_1\rightarrow 0}{W_2(r_m)\over \dot{f}(r_m)}\left(2\dot{W}_1(r_m)-2\sqrt{\dot{W}_1(r_m)\over(l-s_1)}\right),
\ea
with $\cdot$ indicating derivative with respect to $\dt_1$, in which
\ba
\lim_{\dt_1\rightarrow 0}\dot{W}_1(r_m)=\frac{E_{m1}^2 l^4+m_1^2 r_{+}^2 (l+s_1)^2}{m_1^2 r_{+}^2 (l-s_1) (l+s_1)^2},
\ea
and after series expansion, $\dot{f}(r_m)$ becomes:
\ba
\lim_{\dt_1\rightarrow 0}\dot{f}(r_m)=\frac{2 \dt_1  E_{m1}^2 l^4}{m_1^4 r_{+}^2 \left(l^2-s_1^2\right)^2}.
\ea
Eventually, collecting all the above ingredients, $Q$ can be expressed as
\ba
Q&=&\lim_{\dt_1\rightarrow 0}{W_2(r_m)\over \dot{f}(r_m)}\left(2\dot{W}_1(r_m)-2\sqrt{\dot{W}_1(r_m)\over(l-s_1)}\right)=2k\lim_{\dt_1\rightarrow 0}{W_2(r_m)\over \dot{f}(r_m)}\dot{W}_1(r_m)\nn\\
&=&k\lim_{\dt_1\rightarrow 0}W_2(r_m)\frac{m_1^2 (l-s_1) \left(E_{m1}^2 l^4+m_1^2 r_{+}^2 (l+s_1)^2\right)}{ \dt_1  E_{m1}^2 l^4}.\label{tlim}
\ea where $k$ is
\ba
k=1-\sqrt{\dot{W}_1(r_m)\over(l-s_1)}/\dot{W}_1(r_m)=1-\sqrt{\frac{m_1^2 r_{+}^2  (l+s_1)^2}{E_{m1}^2 l^4+m_1^2 r_{+}^2 (l+s_1)^2}}>0.
\ea

Therefore, it is easy to see that the collision center-of-mass energy $E^2_{cm}$ of the two spinning particles diverges because $Q$ diverges at the point $r=r_m$ in the limit $\dt_1\rightarrow 0$. 

\section{Conclusions}\label{conclusion}
In this paper, we have analyzed the collision center-of-mass energy of two spinning particles near BTZ black hole. Our result shows that the center-of-mass
energy of two ingoing spinning particles in the near horizon
limit can be arbitrarily large if one of the particles possesses
a critical angular momentum and the other has a noncritical
angular momentum. However, particle with critical angular momentum can not exist outside of the horizon due to the violation of timelike constraint. Moreover, we proved that the
particle with a subcritical angular momentum is allowed
to exist in near neighbour of an extremal BTZ black hole and the
corresponding collision center-of-mass energy of two spinning particles taking place at the point near
an extremal BTZ black hole can be arbitrarily large in the
$\dt_1\rightarrow 0$ limit.

It should be noted that there are still many important issues that need to be investigated in the future. For example, inspired by the BSW mechanism, people found that the efficiency of
extracting energy from a rotating black hole which is usually called the Penrose process can be greatly improved, especially for spinning particles \cite{M.Bejger2012,Schnittman2014,K.i.Maeda18}. Therefore, with BSW mechanism for spinning particles
in hand, studying the corresponding Penrose process becomes possible, we hope to come to this issue in the near future.

\begin{acknowledgements}

This work is supported by NSFC with No.11775082.

\end{acknowledgements}


\begin{thebibliography}{99}

\bibitem{Piran1975} T. Piran, J. Shaham, and J. Katz, High Efficiency of the Penrose Mechanism for Particle Collisions Astrophys. J. 196, L107
(1975).
\bibitem{banados2009} M. Ba\~nados, J. Silk and S. M. West, Kerr Black Holes as Particle Accelerators to Arbitrarily High Energy, Phys. Rev. Lett. 103, 111102(2009).
\bibitem{banados2011} M. Ba\~nados, B. Hassanain, J. Silk, and S. M. West, Emergent flux from particle collisions near a Kerr black hole, Phys. Rev. D 83, 023004 (2011).
\bibitem{Patil2011} M. Patil and P.S. Joshi, Kerr naked singularities as particle accelerators, Classical Quantum Gravity 28,
235012 (2011).
\bibitem{weishw2010} S. W. Wei, Y. X. Liu, H. Guo, and C. E. Fu, Charged spinning black holes as particle accelerators, Phys. Rev. D 82, 103005 (2010).
\bibitem{liyang2011} Y. Li, J. Yang, Y. Li, S. Wei and Y. Liu Particle acceleration in Kerr-(anti-)de-Sitter black hole backgrounds,  Classical Quantum Gravity 28, 225006 (2011).
\bibitem{zaslav2010}  O. B. Zaslavskii, Acceleration of particles as a universal property of rotating black holes, Phys. Rev. D 82, 083004 (2010).
\bibitem{Abdujabbarov2013}A. Abdujabbarov, N. Dadhich, B. Ahmedov, and H. Eshkuvatov, Particle acceleration around a five-dimensional Kerr black hole, Phys. Rev. D 88, 084036(2013).
\bibitem{Tursunov2013}A. Tursunov, M. Kolos, A. Abdujabbarov, B. Ahmedov, and Z. Stuchlik, Acceleration of particles in spacetimes of black string,  Phys. Rev. D  88, 124001 (2013).

\bibitem{Sadeghi2014} J. Sadeghi, B. Pourhassan, and H. Farahani, Rotating Charged Hairy Black Hole in (2+1) Dimensions and Particle Acceleration, Communications in Theoretical Physics,  Volume 62,  Number 3.
\bibitem{piran1977} T. Piran and J. Shaham,Upper bounds on collisional Penrose processes near rotating black-hole horizons, Phys. Rev. D 16, 1615 (1977).


\bibitem{M.Bejger2012}
M.~Bejger, T.~Piran, M.~Abramowicz, and F.~Hakanson,
Collisional penrose process near the horizon of extreme kerr black holes,
Phys. Rev. Lett. 109, 121101 (2012).
\bibitem{Schnittman2014}
J.~D. Schnittman,
Revised upper limit to energy extraction from a kerr black hole,
Phys. Rev. Lett. 113, 261102 (2014).
\bibitem{berti2015} E. Berti, R. Brito, and V. Cardoso, Ultrahigh-Energy Debris from the Collisional Penrose Process, Phys. Rev. Lett. 114, 251103 (2015).
\bibitem{Leiderschneider} E. Leiderschneider and T. Piran, Maximal efficiency of the collisional Penrose process, Phys. Rev. D 93, 043015
(2016).
\bibitem{Ogasawara2016}K. Ogasawara, T. Harada, and U. Miyamoto, High efficiency of collisional Penrose process requires heavy particle production, Phys. Rev. D
93, 044054 (2016).
\bibitem{Harada2016} T. Harada, K. Ogasawara, and U. Miyamoto, Phys. Rev. D
94, 024038 (2016).
\bibitem{Ogasawara2017}K. Ogasawara, T. Harada, U. Miyamoto, and T. Igata, Escape probability of the super-Penrose process, Phys.
Rev. D 95, 124019 (2017).

\bibitem{K.i.Maeda18}
K.~i. Maeda, K.~Okabayashi, and H.~Okawa,
Maximal efficiency of the collisional penrose process with spinning particles,
Phys. Rev. D 98, 064027 (2018).

\bibitem{btz1992} M. Ba\~nados, C. Teitelboim, and J. Zanelli, Black hole in three-dimensional spacetime, Phys. Rev. Lett. 69, 1849(1992).

\bibitem{tsukamoto2017} N. Tsukamoto, K. Ogasawara, Y. Gong, Particle collision with an arbitrarily high center-of-mass energy near a Ba\~nados-Teitelboim-Zanelli black hole, Phys. Rev. D 96, 024042 (2017).
\bibitem{Mann2009} R. B. Mann, J. J. Oh, and M. I. Park, Role of angular momentum and cosmic censorship in (2+1)-dimensional rotating shell collapse, Phys. Rev. D 79, 064005 (2009).
\bibitem{Rocha2011}J.V. Rocha and V. Cardoso, Gravitational perturbation of the BTZ black hole induced by test particles and weak cosmic censorship in AdS spacetime, Phys. Rev. D 83, 104037(2011).
\bibitem{Tsukamoto2019}Kota Ogasawara, and Naoki Tsukamoto, Effect of the self-gravity of shells on a high energy collision in a rotating Ba\~nados-Teitelboim-Zanelli spacetime ,  Phys. Rev. 99, 024016 (2019).
\bibitem{papapetrou1951} A. Papapetrou, Spinning test-particles in general relativity,
Proc. R. Soc. London A 209, 248 (1951).
\bibitem{dixon1970} W.G. Dixon, Dynamics of extended bodies in general
relativity. I. Momentum and angular momentum, Proc. R.
Soc. A 314, 499 (1970).
\bibitem{wald1972} R. M. Wald, Gravitational spin interaction, Phys. Rev. D 6,
406 (1972).
\bibitem{kerr1963} R. P. Kerr, Gravitational Field of a Spinning Mass as an
Example of Algebraically Special Metrics, Phys. Rev. Lett.
11, 237 (1963).
\bibitem{gaosj2016} M. Guo, S. Gao, Kerr black holes as accelerators of spinning test particles, Phys. Rev. D 93,084025 (2016).
\bibitem{banados2016} C. Armaza, M. Ba\~nados, B. Koch, Collisions of spinning massive particles in a Schwarzschild background, Class. Quantum Grav. 33,105014(2016)
%









\bibitem{zhangyp2016} Y. P. Zhang, B. M. Gu, S. W. Wei, J. Yang, and Y. X. Liu, Charged spinning black holes as accelerators of spinning particles, Phys. Rev. D 94, 124017 (2016).
\bibitem{hackmann2014} E. Hackmann, Cl. L\"ammerzahl, Yu. N. Obukhov, D. Puetzfeld, and I. Schaffer, Motion of spinning test bodies in Kerr spacetime, Phys. Rev. 90, 064035(2014).



\bibitem{saijo1998} M. Saijo, K. Maeda, M. Shibata, and Y. Mino, Gravitational waves from a spinning particle plunging into a Kerr black hole, Phys. Rev. D 58, 064005 (1998).

\bibitem{Jefremov2015} P.I. Jefremov, O. Y. Tsupko, and G. S. Bisnovatyi-Kogan,
Innermost stable circular orbits of spinning test particles in
Schwarzschild and Kerr space-times, Phys. Rev. D 91,
124030 (2015).
\bibitem{jianggao2019}J. Jiang, S. Gao. Universality of BSW mechanism for spinning particles, Eur. Phys. J. C 79:378(2019)










\end{thebibliography}
\end{document}